\documentclass[%
reprint,
showpacs,preprintnumbers,
nofootinbib,
nobibnotes,
amsmath,amssymb,
prb,
floatfix,
longbibliography
]{revtex4-2}

\usepackage{graphicx}
\usepackage{dcolumn}
\usepackage{bm}
\usepackage{colortbl}
\usepackage[unicode=true,colorlinks=true,citecolor=blue,urlcolor=blue]{hyperref}
\usepackage{braket}
\usepackage[normalem]{ulem}
\usepackage{enumitem}

\graphicspath{{img/}}

\renewcommand{\Re}{\mathop{\rm Re}}
\renewcommand{\Im}{\mathop{\rm Im}}

\newcommand{\e}{\mathrm{e}}

\let\ifr\i
\renewcommand{\i}{{\rm i}}

\begin{document}

\title{Proximity effects of chiral magnetization\\ on transition metal dichalcogenide monolayers}

\author{V.~N.~Mantsevich}
\affiliation{Lomonosov Moscow State University, 119991 Moscow, Russia}
\author{I.~S.~Krivenko}
\affiliation{University of Hamburg, 22607 Hamburg, Germany}
\author{D.~S. Smirnov}
\affiliation{Ioffe Institute, 194021 St. Petersburg, Russia}
\email[Electronic address: ]{smirnov@mail.ioffe.ru}

\begin{abstract}
  We consider the proximity effects of a commensurate chiral $120^{\circ}$ N\'eel magnetic structure on the transport and optical properties of a transition metal dichalcogenide monolayer (TMD ML). The enlarged magnetic unit cell leads to the folding of the Brillouin zone and enables efficient intervalley spin-flip scattering. Starting from a six band tight binding model, we develop an effective $k\cdot p$ model, which describes the coupling between charge carriers in the TMD ML and the chiral magnetization. We predict the anomalous antiferromagnetic Hall effect (AHE) for conduction electrons and describe it accounting for the interplay between anomalous velocity, side jump, and skew scattering contributions. The proximity of the chiral magnetization also leads to the mixing and splitting of the exciton resonances in the two valleys which are related by the time reversal symmetry despite zero net magnetization. Finally, we demonstrate the possibility to measure the distribution of in-plane orientation of the chiral magnetization through Faraday rotation and the ellipticity of incident linearly polarized light.
\end{abstract}

\maketitle{}

\section{Introduction}

Magnetic proximity effects provide a way to combine magnetic order with the electrical and optical properties of semiconductors~\cite{zutic_RevModPhys2004,sierra_NatureNanotechnology2021}. These effects are short-ranged and in bulk semiconductors affect only a few atomic layers near the interface. By contrast, in two-dimensional nanostructures, the carrier wave function can lie entirely within the proximity region, so their properties can be strongly modified by proximity effects~\cite{Geim2013,sierra_NatureNanotechnology2021}.

In quantum-well structures, a ferromagnetic film can be deposited close to the quantum well, or magnetic ions can be introduced into the neighboring barrier~\cite{akimov_PhysRevB2017,kalitukha_JChemPhys2023}. The exchange coupling between the magnetic ions and the quantum-well carriers then produces a spin splitting of the electronic states and an equilibrium spin polarization~\cite{akimov_PhysRevB2017,kalitukha_JChemPhys2023}. Spin-dependent carrier capture by the magnetic layer can also produce a large nonequilibrium spin polarization~\cite{korenev_NatCommun2012,Mantsevich_prb_2018,mantsevich_PhysRevB2019}.

Van der Waals heterostructures provide a more direct platform for proximity effects. Atomically thin layers can be placed directly next to magnetic materials without strict lattice-matching conditions and can experience strong proximity effects because of their atomic thickness~\cite{Geim2013,sierra_NatureNanotechnology2021}. In this context, TMD MLs, such as MoS$_2$, MoSe$_2$, WS$_2$, and WSe$_2$, are particularly interesting because they have a direct band gap, strong spin-orbit coupling, and rich valley-dependent physics~\cite{xiao_PhysRevLett2012,cao_NatureCommunications2012,Xu2014}. Magnetic proximity to EuS, for example, can induce spin and valley splittings in these materials, modify exciton resonances, and generate magneto-optical effects~\cite{zhao_NatureNanotech2017,norden_NatCommun2019}. The discovery of atomically thin magnets, including CrI$_3$~\cite{huang_Nature2017}, CrBr$_3$~\cite{Ciorciaro2020}, Cr$_2$Ge$_2$Te$_6$~\cite{Gong2017}, and Fe$_3$GeTe$_2$~\cite{Fei2018}, now allows the fabrication of complex hybrid layered structures with programmable optical and electronic properties~\cite{Burch2018,Gibertini2019,Gong2019}.

Typically, magnetic proximity effects of ferromagnets can be described by an effective exchange magnetic field proportional to the magnetization~\cite{zhao_NatureNanotech2017,Zhong2017,Ciorciaro2020}. For this reason, antiferromagnets are not generally expected to produce strong proximity effects because of their zero net magnetization. In this work, we provide a counterexample to this na{\"{\ifr}}ve expectation.

We consider a TMD ML in proximity to a chiral magnetic $120^{\circ}$ N\'eel structure with a period commensurate with the TMD lattice constant. 
The system under study and its symmetry are described in Sec.~\ref{sec:system}. Then, in Sec.~\ref{sec:model}, 
  we derive an effective $k\cdot p$ description of this hybrid structure from a six-band tight binding model. This effective
model is applied in Secs.~\ref{sec:AHE} and~\ref{sec:optical} to derive the antiferromagnetic AHE and to describe the emerging magneto-optical effects, respectively. Our results are summarized in Sec.~\ref{sec:conclusion}, and some model details are provided in \hyperref[app]{Appendix}.

\section{System and its symmetry}
\label{sec:system}

\begin{figure}
\centering
\includegraphics[width=\linewidth]{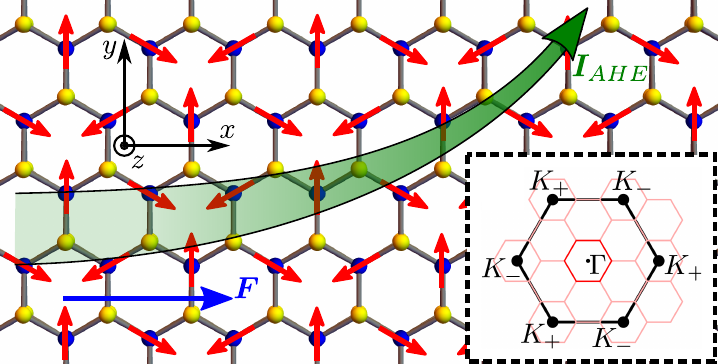}%
\caption{Sketch of the system of a TMD ML covered by a magnetic ML. Blue and yellow spheres show the metal and chalcogen atoms of the TMD ML, respectively. Red arrows denote the direction of the magnetization in the top ML. The blue arrow shows the applied electric field, which leads to the antiferromagnetic AHE (green arrow). The inset shows folding of the Brillouin zone due to the chiral magnetization.} 
\label{fig:symmetry}
\end{figure}

The system under study is sketched in Fig.~\ref{fig:symmetry}. It is based on a TMD ML which is comprised by metal atoms (blue spheres) and chalcogen atoms (yellow spheres). On top of each metal atom we consider magnetic atoms, their spins are shown by red arrows. This in fact means formation of a second magnetic monolayer with the same lattice constant $a_0$ on top of a TMD ML. For dilute magnetic ensembles, the results of our work would remain essentially the same, but the derivations would be more cumbersome. The physical realization of this system is discussed in more detail in Ref.~\onlinecite{i.s.krivenko_jointsubmission2026}, where a few promising physical realizations are mentioned. Unfortunately, at present, no single ideally suitable candidate can be identified without further \textit{ab initio} and experimental studies.

In Ref.~\onlinecite{i.s.krivenko_jointsubmission2026} we have shown that the spin-valley locking of electrons in TMD ML favors a chiral $120^{\circ}$ antiferromagnetic N\'eel state of magnetic impurities, as shown in Fig.~\ref{fig:symmetry}. This structure forms under the assumption of exact commensurability between TMD ML and magnetic lattices. The reason for it is the RKKY interaction mediated by the intervalley scattering of the electrons by magnetic impurities. Its form is the same as for the standard two-dimensional electron gas, but with an additional spin-dependent phase factor related to the spin-valley locking. This factor leads to the effective relative rotations of magnetic moments, which experience the exchange interaction. The period of this rotation equals $3a_0$, as shown in Fig.~\ref{fig:symmetry}. In combination with antiferromagnetic direct exchange interaction between magnetic impurities it leads to the $120^{\circ}$ N\'eel ground state, which simultaneously minimizes the energy of both interactions. The microscopic derivation of the exchange interaction between magnetic impurities~\cite{Mastrogiuseppe2014,Fang2021}, of the magnetic ordering~\cite{i.s.krivenko_jointsubmission2026} and of the magnons is beyond this work. By contrast, we focus on the back action of the chiral magnetization on the electronic and optical properties of the underlying TMD ML.

One can see immediately that the magnetic unit cell in the system is three times larger than the nonmagnetic unit cell. Thus, chiral magnetization leads to the folding of the Brillouin zone as shown in the inset in Fig.~\ref{fig:symmetry}. As a result, the two non-equivalent corners of the Brillouin zone $\bm K_\pm=(\pm4\pi/(3a_0),0)$ become equivalent to the $\Gamma$ point (center of the Brillouin zone). Since the optical properties of a pristine TMD ML are governed by the direct transitions at $K_\pm$ points with the valley-dependent optical selection rules, folding of the Brillouin zone can be expected to hybridize these optical transitions and lead to the splitting of the degenerate optical resonances.

At the microscopic level, folding of the Brillouin zone is caused by the electron scattering by $\bm K_\pm$ on magnetic impurities. This scattering is accompanied by an electron spin flip on the in-plane magnetization. This specific scattering can lead to unusual transport signatures such as antiferromagnetic AHE.

The AHE is described by the components of the conductivity tensor $\sigma_{xy}=-\sigma_{yx}$. According to the Onsager relations, this requires the breaking of time reversal symmetry, so these components are proportional to the odd powers of the magnetization in the absence of an external magnetic field. Intriguingly, the magnetization is absent on average, so one can hardly expect AHE.

However, in systems with the pronounced spin-orbit interaction (such as TMD ML), AHE can emerge even in the absence of net magnetization, and in this case it is called antiferromagnetic AHE~\cite{vsmejkal2022anomalous}. Its possibility is determined by the microscopic symmetry of the system.

The symmetry analysis can be performed in different ways, here we use the simplest considerations suitable specifically for this two dimensional system. First of all, we note that $\sigma_{xy}=-\sigma_{yx}$ transforms in any symmetry group in the same way as the normal component of a magnetic field $B_z$. If the magnetic atoms were placed in the middle of the TMD ML, the system would have a symmetry $\mathcal T\sigma_h$, where $\mathcal T$ denotes time reversal and $\sigma_h$ is the mirror reflection in the horizontal $(xy)$ plane. Since $B_z$ changes sign under this combined symmetry operation, AHE would be forbidden in this case. This shows that for the AHE, it is necessary to distinguish ``top'' and ``bottom'' of the bilayer system.

The in-plane magnetization shown in Fig.~\ref{fig:symmetry} can be rotated by an arbitrary angle $\varphi$ on each atom around the $z$ axis, and this will not significantly change the total energy according to Ref.~\onlinecite{i.s.krivenko_jointsubmission2026}. For the given arbitrary $\varphi$, the point symmetry operations include $E$ (identity) and $2C_3$ (two rotations by $2\pi/3$ around the vertical axis) only; this is group $3$ in Shubnikov notation. Clearly, $B_z$ is invariant in this group, so AHE is allowed in general. In the particular case of $\varphi=\pi k/3$ with integer $k$ (Fig.~\ref{fig:symmetry} corresponds to $\varphi=0$), there are additional symmetry operations $3\mathcal T\sigma_v$ ($\sigma_v$ stands for the vertical reflection plane), the magnetic point symmetry group becomes $3m'$. These operations also leave $B_z$ invariant, so AHE remains allowed. However, in another symmetric case of $\varphi=\pi k/3+\pi/6$ additional operations $3\sigma_v$ appear ($3m$ group), which change the sign of $B_z$ and forbid AHE. This particular case is quite specific and there is no physical reason to favor it, so it seems to be unimportant. Generally, any microscopic calculations of AHE should satisfy these symmetry requirements.

\section{Microscopic model}
\label{sec:model}


\subsection{Tight binding model}

Here, we develop a minimal microscopic model that leads to the optical and electrical effects described above. To do this, we adopt the three band model from Ref.~\onlinecite{PhysRevB.88.085433}, which considers only $d_{0,\pm2}$ orbitals at the metal atoms. The rest of the orbitals and the complete symmetry of the system are taken into account by all possible matrix elements between these orbitals.

Let $\epsilon_1$ and $\epsilon_2$ be the energies of $d_0$ and $d_{\pm 2}$ orbitals. The spin-independent tunneling is described by the following matrix of the tunneling amplitudes between nearest neighbors in the direction $(a_0,0)$ in the basis of ($d_0$, $d_{xy}$ and $d_{x^2-y^2}$) states:
\begin{equation}
  \begin{pmatrix}
    t_0 & t_1 & t_2 \\
    -t_1 & t_{11} & t_{12} \\
    t_2 & -t_{12} & t_{22}
  \end{pmatrix}.
  \label{eq:tun_A}
\end{equation}
The atomic orbitals are related as $d_{\pm2}=(d_{x^2-y^2}\pm\i d_{xy})/\sqrt{2}$. The tunneling amplitudes in the other five directions are given by the rotations of this matrix. In particular, the tunneling matrix in the opposite direction is given by the Hermitian conjugate of Eq.~\eqref{eq:tun_A}.

\begin{figure}
\centering
\includegraphics[width=0.8\linewidth]{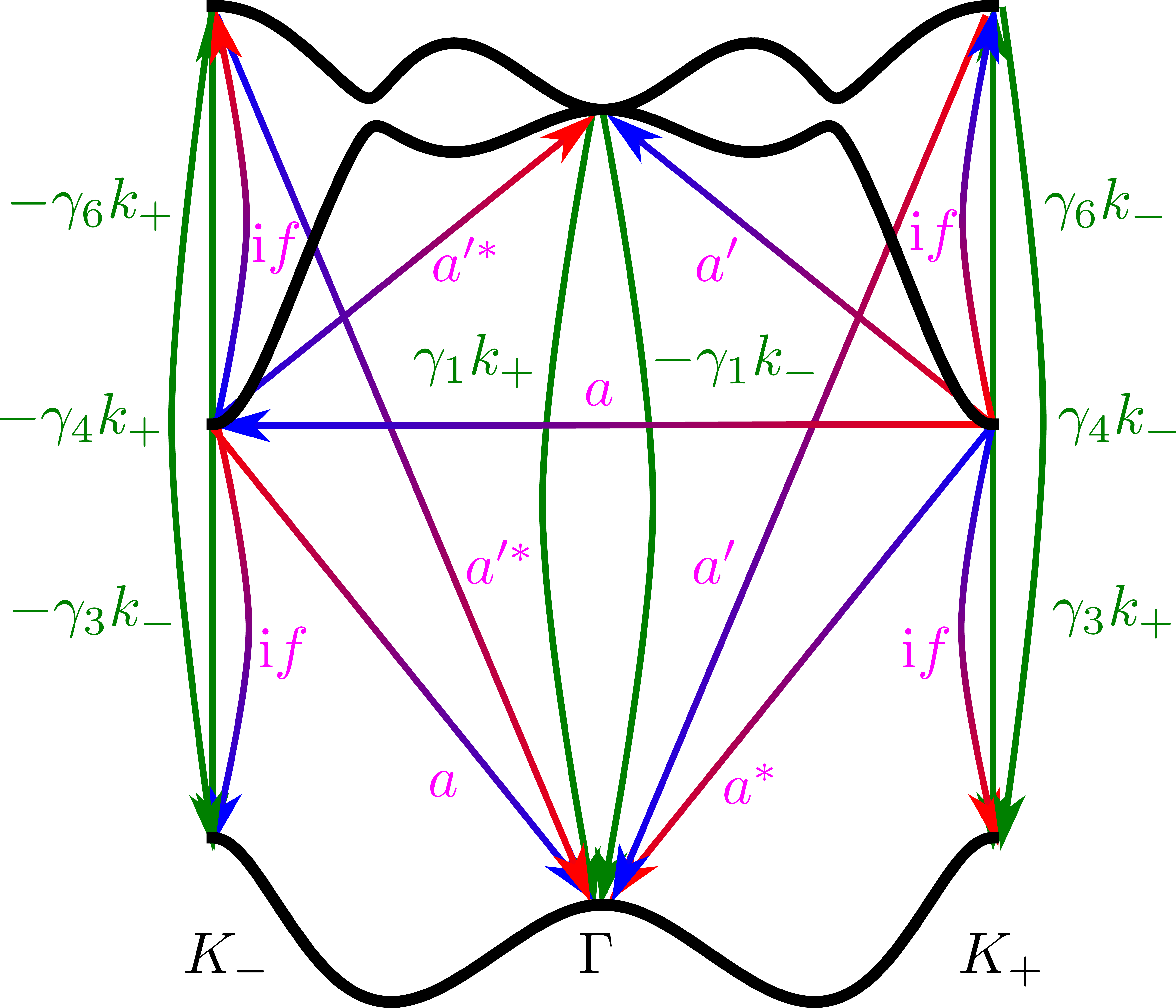}%
\caption{The black solid curves show the energy dispersion of electrons in a pristine TMD ML without spin-orbit interaction calculated for the matrix of tunneling amplitudes~\eqref{eq:tun_A} with the following parameters (in eV)~\cite{PhysRevB.88.085433}: $\epsilon_2-\epsilon_1=1$, $t_0=-0.2$, $t_1=0.4$, $t_2=0.5$, $t_{11}=0.2$, $t_{12}=0.3$, $t_{22}=0.06$. The arrows with labels show matrix elements between the states in $K_\pm$ and $\Gamma$ valleys up to the first order in $\bm k$. The green arrows show spin-independent elements, for other arrows the red and blue ends correspond to spin-up and spin-down states.}
\label{fig:dispersion}
\end{figure}

The corresponding dispersion of electrons in a pristine TMD ML (without spin-orbit interaction) is shown by the black solid lines in Fig.~\ref{fig:dispersion}. Generally, the dispersion consists of three bands: valence band ($v$), conduction band ($c$) and upper conduction band ($c'$). One can see that the valence band has maxima and the conduction band has minima at the $K_\pm$ points, as expected. For the following it is useful to list energies in the $K$ valley:
\begin{subequations}
  \begin{align}
    \begin{split}
      E_{v}(K)&=\epsilon_2-\frac{3}{2}\left(t_{11}+t_{22}\right)-3\sqrt{3}t_{12},\\
      E_c(K)&=\epsilon_1-3t_0,\\
      E_{c'}(K)&=\epsilon_2-\frac{3}{2}\left(t_{11}+t_{22}\right)+3\sqrt{3}t_{12},
    \end{split}
  \end{align}
  and in the $\Gamma$ valley:
  \begin{align}
    \begin{split}
      E_c(\Gamma)&=\epsilon_2+3(t_{11}+t_{22}),\\
      E_v(\Gamma)&=\epsilon_1+6t_0.
    \end{split}
  \end{align}
\end{subequations}

Additional spin-dependent contributions are taken into account by the on-site matrix elements. In the basis of orbitals ($d_0\uparrow$, $d_0\downarrow$, $d_{+2}\uparrow$, $d_{+2}\downarrow$, $d_{-2}\uparrow$, $d_{-2}\downarrow$), the $k$-independent contributions read
\begin{equation}
  \begin{pmatrix}
    0 & AS_- & 0 & -A'S_+ & 0 & -\i f \\
    AS_+ & 0 & -\i f & 0 & -A'S_- & 0 \\
    0 & \i f & \lambda & 0 & 0 & 0 \\
    -A'S_- & 0 & 0 & -\lambda & 0 & 0 \\
    0 & -A'S_+ & 0 & 0 & -\lambda & 0 \\
    \i f & 0 & 0 & 0 & 0 & \lambda
  \end{pmatrix}.
\end{equation}
Here $A$ is the exchange interaction constant between $d_0$ orbitals and spin $\bm S$ of magnetic atoms, $S_\pm=S_x\pm\i S_y$. These spins are considered as classical and frozen. The exchange interaction within $d_{\pm 2}$ orbitals is neglected because of their small extent in the $z$ direction and for simplicity. $A'$ is the constant of weak relativistic exchange, which conserves the angular momentum projection along $z$ axis, but mixes different orbitals, which will be important for the following. We assume both $A$ and $A'$ to be positive while considering magnetization as a classical magnetic moment and neglecting for simplicity its classical and quantum fluctuations~\cite{Chernyshev2009}, which may be particularly important in the vicinity of the phase transitions~\cite{i.s.krivenko_jointsubmission2026}. $\lambda$ is the spin-orbit coupling constant, which is introduced according to Ref.~\onlinecite{PhysRevB.88.085433}. Finally, $f$ takes into account the absence of the horizontal mirror reflection plane by mixing different orbitals and spins in a way that is symmetry equivalent to the normal electric field.

We take into account the magnetic layer perturbatively in the derivation of the effective $k\cdot p$ model.

\subsection{$k\cdot p$ model}

As explained in Sec.~\ref{sec:system}, the chiral magnetization leads to the folding of the Brillouin zone, so $K_\pm$ and $\Gamma$ points become equivalent. The energy dispersion in the vicinity of this point can be described by an effective $k\cdot p$ Hamiltonian, which contains matrix elements of the zeroth and first order in $\bm k$. We obtain this Hamiltonian by calculating the matrix elements of the tight binding Hamiltonian in the basis of the eigenfunctions at the $K_\pm$ and $\Gamma$ points and expanding them in powers of $\bm k$. The resulting Hamiltonian is given explicitly in \hyperref[app]{Appendix}. More clearly it is shown in Fig.~\ref{fig:dispersion} by the arrows. Here, the green arrows show spin-independent matrix elements which are usual for $k\cdot p$ model of a pristine TMD ML. The arrows point from initial to final state, the notations are taken from Ref.~\onlinecite{kormanyos_PhysRevB2013}. From the tight binding model we obtain
\begin{align}
  \begin{split}
    \gamma_1&=\frac{3a_0}{\sqrt{2}}t_1,\quad
    \gamma_3=\frac{3a_0}{2\sqrt{2}}\left(\sqrt{3}t_2+t_1\right),\\
    \gamma_4&=\frac{3\sqrt{3}a_0}{4}\left(t_{22}-t_{11}\right),\quad
    \gamma_6=\frac{3a_0}{2\sqrt{2}}\left(\sqrt{3}t_2-t_1\right).
  \end{split}
\end{align}

The terms related to $\lambda$ are not shown in Fig.~\ref{fig:dispersion}. They lead to the spin splitting of the states $\ket{v,K_\pm}$, $\ket{c',K_\pm}$ and $\ket{c,\Gamma}$ by $2\lambda$~\cite{PhysRevB.88.085433}.

The red-blue arrows show the matrix elements with spin flips caused by the chiral magnetization, their blue and red ends correspond to spin-down and spin-up states, respectively. For them we use the notations $a=\i AS\e^{\i\varphi}$ and $a'=\i A'S\e^{-\i\varphi}$ ($S=|\bm S|$). One can see that the chiral magnetization indeed leads to the scattering by $\bm K_\pm$ with electron spin flips. A few unimportant terms related to $f$ and $a'$ are not shown in the figure. At this point, we would like to note that for the case of a dilute magnetic layer, the constants $a$ and $a'$ should be just reduced proportionally, assuming that the chiral magnetic structure is preserved~\cite{i.s.krivenko_jointsubmission2026}.

Now this compact model can be used for description of electrical and optical properties.

\section{Antiferromagnetic AHE}
\label{sec:AHE}

To describe the modification of transport properties of TMD ML due to the chiral magnetization, we focus on the conduction band, specifically on the $\bm K_\pm$ points of the Brillouin zone, which get coupled. Following the standard quasi-degenerate perturbation theory~\cite{winkler:book}, we account for all other bands in the first order in $\bm k$ and the second order in $\bm S$. This gives the effective Hamiltonian of electrons in the basis of states $|K_+\uparrow\rangle$, $|K_-\uparrow\rangle$, $|K_+\downarrow\rangle$, and $|K_-\downarrow\rangle$:
\begin{equation}
  \label{eq:Hkp}
  \mathcal H_e=
  \begin{pmatrix}
    E_c & \alpha^*k_+ & \i\beta k_- & a^* \\
    \alpha k_- & E_c+\Delta_c & 0 & \i\beta k_- \\
    -\i\beta k_+ & 0 & E_c+\Delta_c & -\alpha^*k_+ \\
    a & -\i\beta k_+ & -\alpha k_- & E_c
  \end{pmatrix}.
\end{equation}
Here, the conduction band spin splitting is
\begin{equation}
  \label{eq:delta}
  \Delta_c=\frac{|a^2|}{E_c(K)-E_v(\Gamma)},
\end{equation}
the Rashba spin-orbit coupling constant is
\begin{equation}
  \beta=f\left[\frac{\gamma_3}{E_c(K)-E_v(K)}+\frac{\gamma_6}{E_{c'}(K)-E_c(K)}\right],
\end{equation}
and $\lambda$ is neglected in comparison with the splittings between different bands. Our numerical calculations show that they have the main effect on $\alpha$ increasing it by approximately 10\% for the realistic parameters. Small unimportant corrections of the second order in the spin-orbit interaction such as $A'^2$, $A'f$, and $f^2$ are omitted as well.

The parameters $\Delta_c$ and $\beta$ are natural to appear and they could be introduced phenomenologically as well. In particular, Rashba spin-orbit coupling is well established in the absence of the horizontal mirror reflection plane~\cite{Zhou2019Rashba,Kezilebieke2020}. The unexpected and important terms in Eq.~\eqref{eq:Hkp} are the couplings between states with the same spin in different valleys, which are determined by
\begin{multline}
  \label{eq:alpha}
  \alpha=\frac{a'}{E_{c}(\Gamma)-E_c(K)}\frac{a^*}{E_c(K)-E_v(\Gamma)}\gamma_1\\
  -\frac{a^*}{E_c(K)-E_v(\Gamma)}\frac{a'}{E_{c'}(K)-E_c(K)}\gamma_6.
\end{multline}
These terms are proportional to the second power of magnetization, vary as $\alpha\propto\e^{-2\i\varphi}$ with its direction, and contain both exchange interaction constants $A$ and $A'$. Note that the derivation of Eq.~\eqref{eq:Hkp} requires coupling strength to the magnetic layer to be smaller than the splittings between different bands and allows for arbitrary relation between all the terms in $\mathcal H_e$ (except for $E_c$).


The simple $4\times4$ conduction band Hamiltonian~\eqref{eq:Hkp} allows a straightforward calculation of the Berry curvature
\begin{equation}
  \label{eq:B_def}
  \mathcal B=2\Im\left\langle\frac{\partial\Psi_{\bm k}}{\partial k_y}\middle|\frac{\partial\Psi_{\bm k}}{\partial k_x}\right\rangle
\end{equation}
of the states, where $\Psi_{\bm k}$ are eigenfunctions at a given wave vector $\bm k$. At $\bm k=0$, for the ground electron state we obtain
\begin{equation}
  \label{eq:B_ans}
  \mathcal B=\frac{4\left|\alpha\right|\beta\cos3\varphi}{(\Delta_c+AS)^2}.
\end{equation}
One can see that it is determined by the direction of magnetization and varies as $\cos3\varphi$. In particular, the Berry curvature vanishes at $3\varphi=\pi/2+\pi k$, exactly when the symmetry analysis of Sec.~\ref{sec:system} forbids AHE.

In an external electric field, the Berry curvature produces an anomalous velocity as $v_y=eE_x\mathcal B/\hbar$, where $e$ is the electron charge. It leads to the anomalous Hall conductivity $\sigma_{xy}=-Ne^2\mathcal B/\hbar$, where $N$ is the density of electrons~\cite{RevModPhys.82.1539}. This contribution is independent of the impurity density, but is generally known to cancel completely with the contribution related to the side jumps that contains the Berry curvature~\cite{PhysRevB.75.045315,Sinitsyn_2008}. The rest of side jumps contribution cancels for short range impurities with the so-called interference part of the skew scattering contribution~\cite{PhysRevB.75.045315,ado_PhysRevB2017,glazov_PhysRevB2020}.

As a result, the anomalous conductivity is determined only by the asymmetric skew scattering contribution. It can be calculated microscopically from the scattering matrix element between $\bm k$ and $\bm k'$ electron states in the lowest conduction subband:
\begin{equation}
  \label{eq:xi_def}
  V_{\bm k',\bm k}=\left\langle\Psi_{\bm k'}\middle|U\delta(\bm r)\middle|\Psi_{\bm k}\right\rangle=U\left[1+\i\frac{\mathcal B}{2}(\bm k\times\bm k')_z\right].
\end{equation}
Here $U$ is the strength of the scattering potential, we use unit sample normalization area, and omit corrections to $\Re V_{\bm k',\bm k}$ quadratic in the wave vector. Taking into account $\Im V_{\bm k',\bm k}$, the standard solution of the Boltzmann kinetic equation for a degenerate electron gas gives the following anomalous Hall conductivity due to the asymmetric phase of the scattering matrix element~\cite{glazov_PhysRevB2020}:
\begin{equation}
  \sigma_{xy}=\pi(Ne)^2\frac{U\tau_p}{\hbar^2}\mathcal B,
\end{equation}
where $\tau_p$ is the momentum relaxation time. The quadratic dependence on $N$ is caused by the additional square of momentum in Eq.~\eqref{eq:xi_def}.

We note that in the realistic case of $AS\gg\Delta_c$, Eq.~\eqref{eq:B_ans} yields the Berry curvature, which is independent of the magnitude of the magnetization $S$, because $\alpha\propto S^2$. However, it depends on the orientation of magnetization as $\propto\cos3\varphi$ in agreement with the symmetry analysis. In the opposite limit of small magnetization, one would have $\sigma_{xy}\propto S^2$.

For the parameters from Fig.~\ref{fig:dispersion}, we obtain an estimate $\mathcal B\sim(A'/A)(\beta/{\rm{eV}})a_0\cos3\varphi$. Using $A'/A=0.1$, $\beta=0.01$~eV$\cdot${\AA}, and $a_0\sim3$~{\AA}, we obtain $\mathcal B\sim3\cdot10^{-3}$~\AA$^2$, which is quite small. To estimate the AHE conductivity, we use electron effective mass $m=0.4m_0$ with $m_0$ being the free electron mass, electron density $N\sim10^{13}$~cm$^{-2}$, momentum relaxation time $\tau_p\sim10$~ps, and $U=10^{-11}$~meV$\cdot$cm$^2$. This gives $\sigma_{xy}\sim10^{-3}e^2/h$ (with $h=2\pi\hbar$). This is a small, but measurable value and comparable to the valley Hall effect caused by electric field, photon and phonon drag in TMD MLs~\cite{glazov_PhysRevB2020}.

\section{Modified optical response}
\label{sec:optical}

Optical properties of TMD MLs around the band edge at low temperatures are determined by tightly bound excitons with small Bohr radius and large binding energy. However, their excitation and recombination are governed by the selection rules for optical transitions between single particle electron and hole states~\cite{MX2Review}. Thus, it is sufficient to consider their modification to describe the effect of the chiral magnetization on the optical properties of the excitons.

From Eq.~\eqref{eq:Hkp} and Fig.~\ref{fig:dispersion} one can see that the most important effect is the mixing of the electron opposite spin states in the $K_+$ and $K_-$ valleys and their splitting by $2|a|$. At $\bm k=0$, the corresponding conduction band states take the form
\begin{equation}
  \ket{\Psi_\pm}=\frac{1}{\sqrt{2}}\left(\ket{c,K_+,\uparrow}\pm\i\e^{\i\varphi}\ket{c,K_-,\downarrow}\right).
\end{equation}
As a result, the intervalley and intravalley excitons hybridize, which leads to the splitting of the exciton resonance into two. Note that this splitting is not modified by the Rashba spin splitting derived above to the first order, because it is odd in $\bm k$. The second order corrections are small (on the order of 1\%) and can be neglected.

\begin{figure}
\centering
\includegraphics[width=0.7\linewidth]{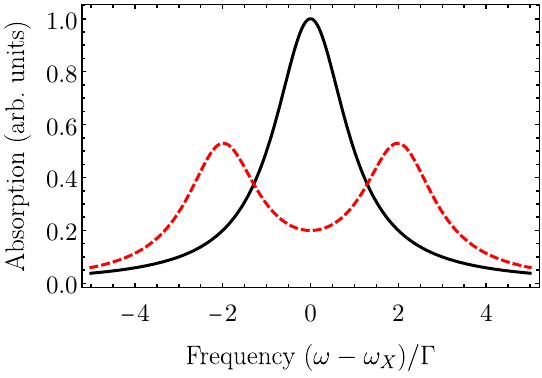}%
\caption{Illustration of the absorption spectrum of a pure TMD ML in the vicinity of an exciton resonance frequency $\omega_X$ (black solid line) and its modification due to the chiral magnetization (red dashed line).}
\label{fig:optical}
\end{figure}

The splitting of the exciton resonance is illustrated in Fig.~\ref{fig:optical}, where we take into account that the two resonances share the transition oscillator strength, so they are two times weaker than the initial resonance. For this figure, we consider a phenomenological broadening $\Gamma=|a|/(2\hbar)$, which we introduce here for illustration and consider as homogeneous. In real structures, there can be a significant inhomogeneous broadening due to random strain, variations of dielectric screening, charged impurities, structural defects, etc~\cite{raja_NatNanotechnol2019,shao_PhysRevMaterials2022}, but it does not change this figure qualitatively. Importantly, the splitting $2|a|$ can be as large as tens of millielectronvolts in the case of strong exchange interaction and significantly exceed width of the exciton resonance. At the same time, the optical transitions in the two circular polarizations remain degenerate and independent.

Curiously, in addition to this, the chiral magnetization folds the opposite valleys to the same point of the reduced Brillouin zone, as explained above, and this leads to the brightening of the intervalley transitions. Specifically, from Fig.~\ref{fig:dispersion}, one can see that in the second order of perturbation theory the conduction band states acquire the following corrections:
\begin{align}
  \begin{split}
    \ket{c,K_+,\downarrow}&-\frac{a^*}{E_c(K)-E_v(\Gamma)}\frac{a'}{E_{c'}(K)-E_{c}(K)}\ket{c',K_-,\downarrow},\\
    \ket{c,K_-,\uparrow}&-\frac{a}{E_c(K)-E_v(\Gamma)}\frac{a'^*}{E_{c'}(K)-E_{c}(K)}\ket{c',K_+,\uparrow}.
  \end{split}
\end{align}
This leads to the optical transitions in $\sigma^\pm$ polarizations, respectively due to the $\gamma_4$ matrix element of $k\cdot p$ effective Hamiltonian. Using the parameters from Fig.~\ref{fig:dispersion}, $|a|=50$~meV and $|a'|=5$~meV, we find that the corresponding matrix elements are approximately five orders of magnitude smaller than those for the main transitions (related to $\gamma_3$ matrix element), so the intervalley transitions are unlikely to be observed experimentally.

A very sensitive experimental optical technique is the measurement of Faraday rotation and ellipticity of linearly polarized light, which is transmitted through or reflected from the sample. These effects have the same symmetry as the AHE, i.e. transform as $B_z$.

To describe them microscopically, we consider corrections to the conduction band states, which follow from the effective $k\cdot p$ Hamiltonian, as one can see from Fig.~\ref{fig:dispersion}:
\begin{align}
  \label{eq:K_dark}
  \begin{split}
    \left|c,K_+,\uparrow\right>&-\frac{a'}{E_c(\Gamma)-E_c(K)}\left|c,\Gamma,\downarrow\right>,\\
    \left|c,K_-,\downarrow\right>&-\frac{a'^*}{E_c(\Gamma)-E_c(K)}\left|c,\Gamma,\uparrow\right>.
  \end{split}
\end{align}
Similar corrections to the valence band states have the form
\begin{align}
  \begin{split}
    \left|v,K_-,\downarrow\right>&+\frac{\i f}{E_c(K)-E_v(K)}\frac{a}{E_v(K)-E_v(\Gamma)}\left|v,\Gamma,\downarrow\right>,\\
    \left|v,K_+,\uparrow\right>&+\frac{\i f}{E_c(K)-E_v(K)}\frac{a^*}{E_v(K)-E_v(\Gamma)}\left|v,\Gamma,\uparrow\right>.
  \end{split}
\end{align}
Due to these corrections, the optical transition matrix elements for transitions to the ground electron state $\Psi_-$ in the $\sigma^\pm$ polarizations take the form
\begin{equation}
  \label{eq:Mpm}
  M_\pm\propto\pm 1+\delta\e^{\pm3\i\varphi},
\end{equation}
where
\begin{equation}
  \delta=\frac{f}{E_c(K)-E_v(K)}\frac{|a|}{E_v(K)-E_v(\Gamma)}\frac{|a'|}{E_c(\Gamma)-E_c(K)}\frac{\gamma_1}{\gamma_3}.
\end{equation}
This shows that the absolute values of the matrix elements in the two circular polarizations can be different. 

For excitons, the optical transition matrix elements should be averaged over the wave functions. For estimations, we assume that the optical transition matrix elements in the form of Eq.~\eqref{eq:Mpm} can be applied for the finite wave vectors up to $k\sim 1/a_B$ with $a_B$ being the exciton Bohr radius. Then the Faraday rotation ($\theta_F$) and ellipticity ($\theta_E$) of the transmitted linearly polarized light can be estimated as~\cite{glazov:review}
\begin{equation}
  \label{eq:thetas}
  \theta_F+\i\theta_E\sim\frac{\mathcal A\Gamma}{\omega-\omega_X+\i\Gamma}\delta\cos3\varphi,
\end{equation}
where $\mathcal A$ is the absorbance at the exciton resonance frequency $\omega_X$ in a pristine TMD ML. The Faraday rotation and ellipticity in Eq.~\eqref{eq:thetas} are described by the dispersive and absorptive lineshapes as usual. One can see that $\theta_{F,E}$ scale as $\delta\propto aa'$, i.e. quadratically with exchange interaction. For the parameters used above and $\mathcal A\sim0.1$, we obtain an estimate $\theta_{F,E}\sim10^{-7}$ (in radians), which is small, but a typical value, for example, for the spin-noise measurements~\cite{Smirnov-review}.

An important feature of the Faraday rotation in this case is that it is produced by the in-plane chiral magnetization, which has no component normal to the TMD ML. From Eq.~\eqref{eq:thetas} one can also see that these effects depend on the orientation of this magnetization as $\cos3\varphi$. Thus, in principle, it is possible to measure the spatial distribution of the magnetization orientation in the samples.



\section{Conclusion}
\label{sec:conclusion}

We have described the proximity effects of a chiral two dimensional $120^{\circ}$ N\'eel state on a TMD ML. The symmetry analysis reveals the possibility of an antiferromagnetic AHE and a splitting of the optical exciton resonances in this structure. These effects are described on the basis of the original tight binding model and an effective $k\cdot p$ model derived from it. The calculated AHE conductivity is measurable, while the splitting of the exciton resonances can be large. In addition, we have demonstrated that linearly polarized light incident on this structure experiences Faraday rotation and acquires a finite ellipticity. All these effects demonstrate that the proximity of an antiferromagnetic ML to a TMD ML can drastically modify its electrical and optical properties even in the absence of any net magnetization.


\section{Acknowledgments}

We thank A.V. Shumilin for fruitful discussions. V.N.M. thank for support Russian Science Foundation grant No. 25-12-00093. 
The derivation of the effective $k\cdot p$ model by D.S.S. was supported by the Russian Science Foundation Grant No. 25-72-10031.

\appendix*
\section{Effective $k\cdot p$ Hamiltonian}
\label{app}

Explicitly, the effective $k\cdot p$ Hamiltonian shown in Fig.~\ref{fig:dispersion} can be written in the basis of states
$\ket{c',K_+,\uparrow}$, $\ket{c',K_-,\uparrow}$, $\ket{c',K_+,\downarrow}$, $\ket{c',K_-,\downarrow}$,
$\ket{c,K_+,\uparrow}$, $\ket{c,K_-,\uparrow}$, $\ket{c,K_+,\downarrow}$, $\ket{c,K_-,\downarrow}$,
$\ket{v,K_+,\uparrow}$, $\ket{v,K_-,\uparrow}$, $\ket{v,K_+,\downarrow}$, $\ket{v,K_-,\downarrow}$,
$\ket{c',\Gamma,\uparrow}$, $\ket{c',\Gamma,\downarrow}$, $\ket{c,\Gamma,\uparrow}$, $\ket{c,\Gamma,\downarrow}$, $\ket{v,\Gamma,\uparrow}$, $\ket{v,\Gamma,\downarrow}$. It is useful to divide the Hamiltonian into blocks as follows:
\begin{equation}
  \mathcal H_{k\cdot p}=
  \begin{pmatrix}
    \mathcal H_{c',c'} & \mathcal H_{c',c} & \mathcal H_{c',v} & \mathcal H_{c',\Gamma} \\
    \mathcal H_{c',c}^{\dagger} & \mathcal H_{c,c} & \mathcal H_{c,v} & \mathcal H_{c,\Gamma} \\
    \mathcal H_{c',v}^{\dagger} & \mathcal H_{c,v}^{\dagger} & \mathcal H_{v,v} & 0 \\
    \mathcal H_{c',\Gamma}^{\dagger} & \mathcal H_{c,\Gamma}^{\dagger} & 0 & \mathcal H_{\Gamma,\Gamma}
  \end{pmatrix}.
\end{equation}
Here, the subscripts $c'$, $c$, and $v$ refer to the states in the $K_\pm$ valleys in the corresponding bands, and the subscript $\Gamma$ refers to the states in the $\Gamma$ valley. The diagonal blocks have the form
\begin{widetext}
  \begin{equation}
    \mathcal H_{c',c'} =
    \begin{pmatrix}
      E_{c'}(K)-\lambda & 0 & 0 & 0 \\
      0 & E_{c'}(K)+\lambda & 0 & 0 \\
      0 & 0 & E_{c'}(K)+\lambda & 0 \\
      0 & 0 & 0 & E_{c'}(K)-\lambda
    \end{pmatrix},
    \qquad
    \mathcal H_{c,c} =
    \begin{pmatrix}
      E_c(K) & 0 & 0 & a^* \\
      0 & E_c(K) & 0 & 0 \\
      0 & 0 & E_c(K) & 0 \\
      a & 0 & 0 & E_c(K)
    \end{pmatrix},
    \nonumber
  \end{equation}
  \begin{equation}
    \mathcal H_{v,v} =
    \begin{pmatrix}
      E_v(K)+\lambda & 0 & 0 & 0 \\
      0 & E_v(K)-\lambda & 0 & 0 \\
      0 & 0 & E_v(K)-\lambda & 0 \\
      0 & 0 & 0 & E_v(K)+\lambda
    \end{pmatrix},
  \end{equation}
  \begin{equation}
    \mathcal H_{\Gamma,\Gamma} =
    \begin{pmatrix}
      E_c(\Gamma)+\lambda & 0 & 0 & 0 & -\gamma_1 k_+ & \i f \\
      0 & E_c(\Gamma)+\lambda & 0 & 0 & \i f & \gamma_1 k_- \\
      0 & 0 & E_c(\Gamma)-\lambda & 0 & \gamma_1 k_- & 0 \\
      0 & 0 & 0 & E_c(\Gamma)-\lambda & 0 & -\gamma_1 k_+ \\
      -\gamma_1 k_- & -\i f & \gamma_1 k_+ & 0 & E_v(\Gamma) & 0 \\
      -\i f & \gamma_1 k_+ & 0 & -\gamma_1 k_- & 0 & E_v(\Gamma)
    \end{pmatrix}.
    \nonumber
  \end{equation}
\end{widetext}
The off-diagonal blocks read:
\begin{subequations}
  \begin{equation}
    \mathcal H_{c',c} =
    \begin{pmatrix}
      \gamma_6 k_- & 0 & 0 & 0 \\
      0 & -\gamma_6 k_+ & 0 & \i f \\
      \i f & 0 & \gamma_6 k_- & 0 \\
      0 & 0 & 0 & -\gamma_6 k_+
    \end{pmatrix},
  \end{equation}
  \begin{equation}
    \mathcal H_{c',v} =
    \begin{pmatrix}
      \gamma_4 k_+ & 0 & 0 & 0 \\
      0 & -\gamma_4 k_- & 0 & 0 \\
      0 & 0 & \gamma_4 k_+ & 0 \\
      0 & 0 & 0 & -\gamma_4 k_-
    \end{pmatrix},
  \end{equation}
  \begin{equation}
    \mathcal H_{c',\Gamma} =
    \begin{pmatrix}
      0 & 0 & 0 & 0 & 0 & a'^* \\
      0 & 0 & 0 & 0 & 0 & 0 \\
      0 & 0 & 0 & 0 & 0 & 0 \\
      0 & 0 & 0 & 0 & a' & 0
    \end{pmatrix},
  \end{equation}
  \begin{equation}
    \mathcal H_{c,v} =
    \begin{pmatrix}
      \gamma_3 k_- & 0 & 0 & 0 \\
      0 & -\gamma_3 k_+ & a'^* & -\i f \\
      -\i f & a' & \gamma_3 k_- & 0 \\
      0 & 0 & 0 & -\gamma_3 k_+
    \end{pmatrix},
  \end{equation}
  \begin{equation}
    \mathcal H_{c,\Gamma} =
    \begin{pmatrix}
      0 & 0 & 0 & a'^* & 0 & 0 \\
      0 & 0 & 0 & 0 & 0 & a^* \\
      0 & 0 & 0 & 0 & a & 0 \\
      0 & 0 & a' & 0 & 0 & 0
    \end{pmatrix}.
  \end{equation}
\end{subequations}

\renewcommand{\i}{\ifr}
%

\end{document}